\documentclass[12pt,onecolumn,twoside]{CTT}

\usepackage{graphicx}

\usepackage{multicol}

\usepackage{amsmath}
\usepackage{amsthm}
\usepackage{amsfonts}
\usepackage{amssymb}
\usepackage{mathrsfs}

\usepackage{lettrine}

\usepackage[sort]{cite}

\usepackage{times}

\usepackage{booktabs}

\usepackage{wasysym}

\usepackage{fancyhdr}

\usepackage{ulem}

\usepackage{picins}

\def\ubar#1{\underset{\raise0.3em\hbox{$\smash{\scriptscriptstyle-}$}}{#1}}

\allowdisplaybreaks

\newtheoremstyle{mystyle}{1pt}{1pt}{\normalfont}{\parindent}{\bfseries}{}{1em}{}
\theoremstyle{mystyle}

\renewcommand{\footnotesep}{1pt}
\renewcommand{\footnotesize}{\scriptsize}


\newcommand{\R}{\mathbb{R}}
\newcommand{\nn}{\mathsf{n}}

\newcommand{\bb}{\mathsf{b}}
\newcommand{\aaa}{\mathsf{a}}
\newcommand{\fff}{\mathsf{f}}

\newcommand{\ppp}{\mathsf{p}}
\newcommand{\ccc}{\mathsf{c}}

\newcommand{\xxx}{\mathsf{x}}

\newcommand{\AAA}{\mathsf{A}}

\newcommand{\EEE}{\mathsf{E}}

\newcommand{\TTT}{\mathsf{T}}
\newcommand{\ddd}{\mathsf{d}}
\newcommand{\pp}{\mathfrak{p}}

\begin{document}                          
\makeatletter
\def\@autr{{H.~Shim} et al.}             
\makeatother

\begin{frontmatter}                       

\title{Yet Another Tutorial of Disturbance Observer: Robust Stabilization and Recovery of Nominal Performance}\footnotetext
{\scriptsize $^{\dag}$Corresponding author. E-mail: hshim@snu.ac.kr. Tel.: +82-2-880-1745.
This work was supported partially by University of Florida, USA, and by the National Research Foundation of Korea (NRF) grant funded by the Korea government (MISP) (2015R1A2A2A01003878).}

\author[1]{Hyungbo SHIM}$^{\dag}$,{ }
\author[1]{Gyunghoon PARK},{ }
\author[2]{Youngjun JOO},{ }
\author[3]{Juhoon BACK},{ }
\author[4]{Nam Hoon JO}

\address[1]{. ASRI, Department of Electrical and Computer Engineering, Seoul National University, Korea;}
\address[2]{. Department of Electrical Engineering and Computer Science, University of Central Florida, USA;}
\address[3]{. School of Robotics, Kwangwoon University, Korea;}
\address[4]{. Department of Electrical Engineering, Soongsil University, Korea}

\begin{abstract} 
	This paper presents a tutorial-style review on the recent results about the disturbance observer (DOB) in view of robust stabilization and recovery of the nominal performance.
	The analysis is based on the case when the bandwidth of Q-filter is large, and it is explained in a pedagogical manner that, even in the presence of plant uncertainties and disturbances, the behavior of real uncertain plant can be made almost similar to that of disturbance-free nominal system both in the transient and in the steady-state.
	The conventional DOB is interpreted in a new perspective, and its restrictions and extensions are discussed.
\end{abstract}

\begin{keyword}
	Disturbance observer, robust stabilization, robust transient response, disturbance rejection.
\end{keyword}

\end{frontmatter}

\makeatletter
\renewcommand{\headrulewidth}{0pt}
\fancypagestyle{plain}{\fancyhf{} \fancyhead[LE,LO]{}
\fancyhead[CE,CO]{} \fancyhead[RE,RO]{} \lfoot{} \cfoot{}
\rfoot{}} \thispagestyle{plain}
\pagestyle{fancy}\fancyhf{} \fancyhead[RE]{} \fancyhead[CE]{} \fancyhead[LE,RO]{\thepage}
\fancyhead[CO]{}
\fancyhead[LO]{}
\fancyfoot[LE,RO]{}\fancyfoot[RE,LO]{}
\fancyfoot[CE,CO]{} \makeatother

\begin{multicols}{2}                

\section{Introduction}

Robust control via disturbance observer (DOB) has many advantages over other robust control methods.
In particular, it is an inner-loop controller and its primary role is just to compensate uncertainty in the plant and external disturbances into the plant, so that the inner-loop behaves like a nominal plant without disturbances and uncertainties.
Therefore, any outer-loop controller that is designed just for the nominal plant without considering robustness should work, and this enables modular design of controllers; that is, the outer-loop controller deals with nominal stability and nominal performance, and the inner-loop DOB cares for robustness against uncertainty and disturbances.
In this sense, DOB is in contrast to other robust control methods such as $\mathcal{H}_\infty$ control, adaptive control, or sliding mode control, and there is much design freedom for the outer-loop controller in DOB-based robust control.
When there is no uncertainty and disturbance, the DOB-based robust control shows the best nominal performance without intervention of the inner-loop DOB, while the performance degrades gradually as the amount of uncertainty and disturbance grows.
Finally, DOB has the benefit of design simplicity (while its theoretical analysis is not simple) so that it has been employed in many industrial applications.

Because of the benefits, a large number of research works have been reported in the literature, including survey-style papers \cite{GC05,SO14,CYGS15}, monographs \cite{LYCWC14,GC13}, and a related paper \cite{TG09,G15,Gao14,Xue15} under the name of `active disturbance rejection control (ADRC).'
On the other hand, this paper presents yet another tutorial of DOB as a summary of recent findings by the authors, in less formal style (for example, we avoid the theorem-proof style of writing).
We view the DOB\footnote{The origin of DOB, which was called a load torque estimator, dates back to \cite{OOM83}. It was more or less an estimator rather than a robust controller.} as an output feedback robust controller which, under certain conditions such as minimum-phaseness of the plant and large bandwidth of the Q-filters, enables robust stabilization against arbitrarily large parametric uncertainty (as long as the uncertain parameters are bounded and their bounds are known a priori), and recovery of nominal steady-state and transient performance.
This perspective will lead us to the underlying principles of the DOB that has large bandwidth of Q-filters.

\section{System Description for Analysis}

The systems dealt with in this paper are the single-input-single-output linear time-invariant systems given by
\begin{equation}\label{eq:origsys}
	\begin{aligned}
		\dot \xxx &= \AAA \xxx + \bb u + \EEE \ddd, \quad & \xxx &\in \R^n, \; u \in \R, \\
		y &= \ccc \xxx, & \ddd &\in \R^q, \; y \in \R, 
	\end{aligned}
\end{equation}
where $u$ is the input, $y$ is the output, $\xxx$ is the state, and $\ddd$ is the external disturbance.
The disturbance signal $\ddd(t)$ is assumed to be smooth (i.e., differentiable as many times as necessary with respect to time $t$), and we assume that $\ddd(t)$ and its derivatives are uniformly bounded.
The matrices $\AAA$, $\bb$, $\ccc$, and $\EEE$ are of appropriate sizes, and are assumed to be uncertain.
In particular, we assume that system \eqref{eq:origsys} (without the disturbance term $\EEE\ddd$) is a minimal realization of the transfer function
\begin{equation}\label{eq:P(s)}
	P(s) = \frac{\beta_m s^m + \beta_{m-1} s^{m-1} + \cdots + \beta_0}{s^n + \alpha_{n-1} s^{n-1} + \cdots + \alpha_0} = \ccc (sI-\AAA)^{-1}\bb
\end{equation}
in which, all parameters $\alpha_i$ and $\beta_i$ are uncertain, but $\beta_m \not = 0$ and the sign of $\beta_m$ (which is so-called the high-frequency gain of $P(s)$) is known.
System \eqref{eq:origsys} can always be transformed to
\begin{align}\label{eq:sys}
	\begin{split}
		y &= x_1 \\
		\dot x_1 &= x_2 \\
		& \vdots \\
		\dot x_{\nu-1} &= x_\nu \\
		\dot x_\nu &= \phi x + \psi z + g u + g d =: f(x,z) + gu + gd \\
		\dot z &= Sz + Gx + d_z =: h(x,z,d_z)
	\end{split}
\end{align}
where $\nu := n-m$, $x=[x_1,\cdots,x_\nu]^T \in \R^\nu$, and $z \in \R^m$.
Here, the notation $f$ and $h$ are defined for convenience, which will be used frequently later.
The disturbance signals $d$ and $d_z$ are linear combinations of $\ddd$ and its derivatives (for details, refer to the Appendix, e.g., equation \eqref{eq:ddd}).
All the matrices $\phi \in \R^{1 \times \nu}$, $\psi \in \R^{1 \times m}$, $g \in \R^{1 \times 1}$, $S \in \R^{m \times m}$, and $G \in \R^{m \times \nu}$ are uncertain, but the sign of $g$ is known and $g \not = 0$ (this is because $g$ is in fact $\beta_m$, which is clarified in the Appendix).
Refer to the Appendix for the derivation from \eqref{eq:origsys} to \eqref{eq:sys}.
If the plant has the input disturbance only (like in Fig.~\ref{fig:CDOB}), then the plant can be written as in \eqref{eq:sys} without the term $d_z$ in the $z$-subsystem.

The representation \eqref{eq:sys} is called the {\it normal form} \cite{Isidori95,Khalil02}.\footnote{
	It is the name of the structure like the well-known controllability/observability canonical form of the plant.
	The representation \eqref{eq:sys} of the transfer function \eqref{eq:P(s)} can also be directly derived. 
	See \cite[p.~513--514]{Khalil02} for this procedure.
	By this procedure, it is also seen that, in a certain coordinate, the term $Gx$ in \eqref{eq:sys} can depend only on $x_1$, so that, $Gx$ can be written like $Gx = G' x_1 = G' y$.}
The reason for writing the system state in the split form of $x$ and $z$ is to emphasize their different roles that will be seen shortly.
The integer $\nu$ is called the {\it relative degree} of the plant.
It is emphasized that the eigenvalues of the matrix $S$ are the zeros of $P(s)$ in \eqref{eq:P(s)} (see the Appendix for the proof), and $\dot z = Sz$ is called the {\it zero dynamics} of the system.
Then, we say the system is of {\it minimum phase} if and only if the matrix $S$ is Hurwitz.
For designing the DOB, it is not necessary to convert the given plant into the normal form.
The representation \eqref{eq:sys} is just for the analysis in this paper.

\section{Required Action for DOB}

The behavior of \eqref{eq:sys} is unexpected because the plant $P(s)$ is uncertain, and thus, the quantities $\phi$, $\psi$, $g$, $G$, and $S$ are consequently uncertain.
Hence, one may want to design a control input $u$ such that the system \eqref{eq:sys} behaves like its nominal plant:
\begin{align}\label{eq:nominal}
	\begin{split}
		y = x_1, \qquad 
		\dot x_i &= x_{i+1}, \quad i = 1,\dots,\nu-1, \\
		\dot x_\nu &= f_\nn(x,z) + g_\nn \bar u , \\
		\dot z &= h_\nn(x,z)
	\end{split}
\end{align}
where $\bar u$ is an external input that is designed by another (outer-loop) controller.
Comparing \eqref{eq:sys} and \eqref{eq:nominal}, it is seen that system \eqref{eq:nominal} is the system \eqref{eq:sys}, in which, there is no disturbance and the uncertain $f$, $g$, and $h$ are replaced with their nominal $f_\nn$, $g_\nn$, and $h_\nn$, respectively, where $f_\nn(x,z) = \phi_\nn x + \psi_\nn z$ and $h_\nn(x,z) = S_\nn z + G_\nn x$.
While the replacement of $f$ and $g$ can be achieved if the control $u$ in \eqref{eq:sys} becomes the same as
\begin{equation*}
	u_{\rm candidate} = -d + \frac{1}{g}\left( - f(x,z) + f_\nn(x,z) \right) + \frac{g_\nn}{g} \bar u ,
\end{equation*}
the replacement of $h$ is a difficult task because the $z$-subsystem is not directly affected by the control $u$.
Hence, instead of replacing $h$ in the $z$-subsystem of \eqref{eq:sys}, let us introduce a new state $z_\nn \in \R^m$ (which will be implemented in the controller) and construct a new (dynamic) desired input as
\begin{align}\label{eq:desired}
	\begin{split}
		\dot z_\nn &= h_\nn(x,z_\nn),  \\
		u_{\rm desired} &= -d + \frac{1}{g}\left( - f(x,z) + f_\nn(x,z_\nn) \right) + \frac{g_\nn}{g} \bar u . 
	\end{split}
\end{align}
Now if $u \equiv u_{\rm desired}$ (for all $t \ge 0$), then the system \eqref{eq:sys} becomes
\begin{align}
	y = x_1, \qquad 
	\dot x_i &= x_{i+1}, \quad i = 1,\dots,\nu-1, \label{eq:clsys1} \\
	\dot x_\nu &= f_\nn(x,z_\nn) + g_\nn \bar u , \label{eq:clsys2} \\
	\dot z_\nn &= h_\nn(x,z_\nn), \label{eq:clsys3} \\
	\dot z &= h(x,z,d_z). \label{eq:clsys4} 
\end{align}
Clearly, the system \eqref{eq:clsys1}--\eqref{eq:clsys3} yields the same behavior as \eqref{eq:nominal}.
At the same time, the $z$-subsystem \eqref{eq:clsys4} becomes stand-alone and does not affect the output $y$.
In other words, the state $z$, which was observable from $y$ in \eqref{eq:sys}, has now become unobservable by the desired input $u_{\rm desired}$.\footnote{This is not possible in practice because $u_{\rm desired}$ contains unknown quantities and so we cannot let $u \equiv u_{\rm desired}$. However, since the DOB will estimate $u_{\rm desired}$ and let $u \approx u_{\rm desired}$, the degree of observability of $z$ at least gets weakened.}
This is the cost to pay for enforcing the nominal input-output behavior of \eqref{eq:nominal}, or \eqref{eq:clsys1}--\eqref{eq:clsys3}, upon the real plant \eqref{eq:sys}.
Since we do not have any information of $z$ from the output $y$ (when this nominal input-output behavior is achieved) and have no more freedom left in the input $u$ ($=u_{\rm desired}$) to control $z$-subsystem, we have to ask that $z$-subsystem is stable itself (i.e., $S$ is Hurwitz) so that the state $z(t)$ does not diverge under bounded $x$ and $d_z$.

In order to implement the control idea discussed so far, there are still two more challenges.
First, to implement \eqref{eq:desired}, the state $x$ needs to be estimated because $x$ is not directly measured but is used to compute the nominal values of $f_\nn(x,z_\nn)$ and $h_\nn(x,z_\nn)$.
This problem may be solved by a state observer, but a {\it robust} estimation of $x$ is necessary since the system \eqref{eq:sys} is uncertain and is affected by disturbances.
Second, since $u_{\rm desired}$ contains unknown quantities such as $d$, $f(x,z)$, and $g$, we cannot compute it directly.
Instead, we have to estimate $u_{\rm desired}(t)$ and drive $u(t)$ to the estimate.

	\begin{center}
		\includegraphics[width=0.4\textwidth]{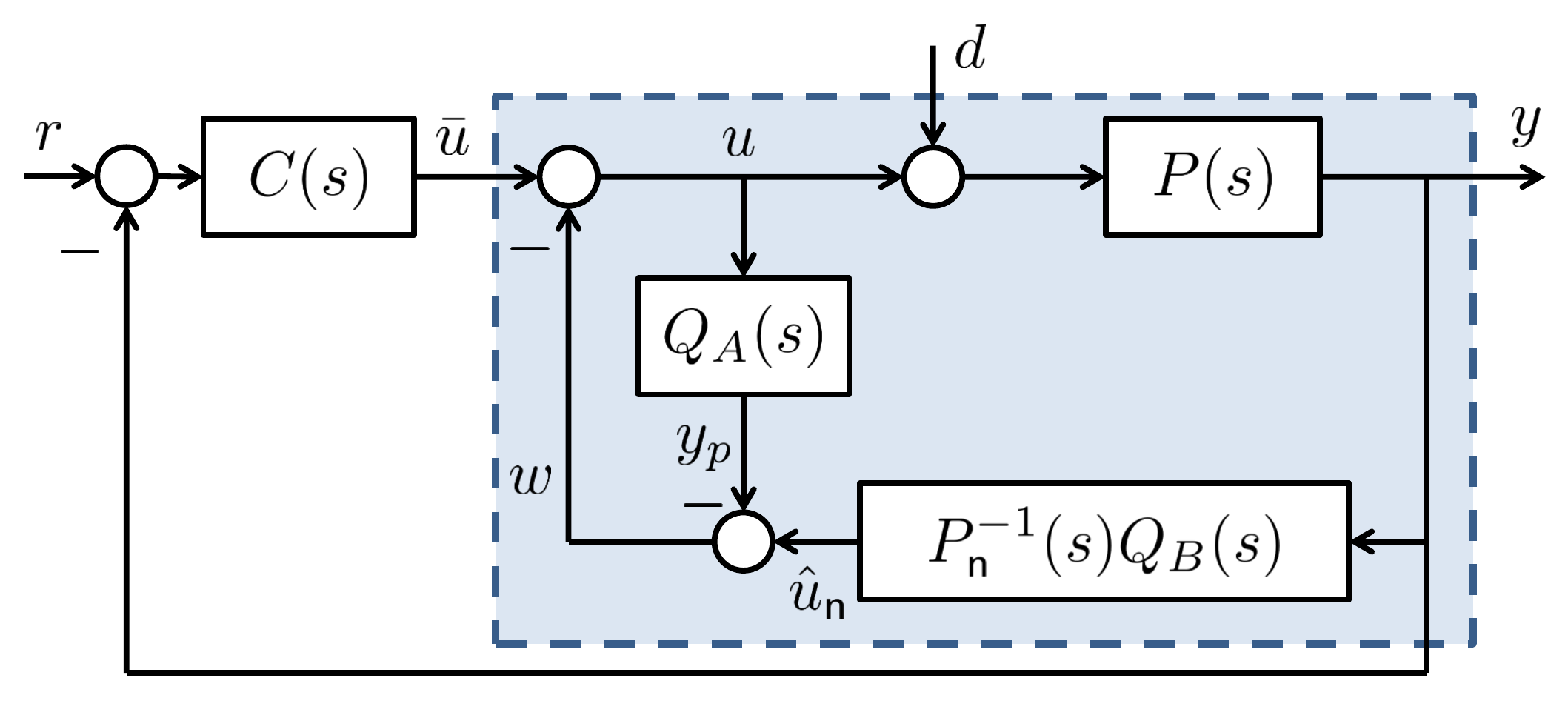}
		\figurecaption{Closed-loop system with DOB structure; $r$ is a reference signal, and $d$ is an input disturbance. The DOB can also be combined with $C(s)$ when implemented, which becomes then a feedback controller having two inputs $r$ and $y$.}\label{fig:CDOB}
	\end{center}

It may be rather surprising and exciting to see that the conventional disturbance observer \cite{KT13,SO14,CYGS15}, depicted in Fig.~\ref{fig:CDOB}, is performing all the afore-mentioned tasks when the bandwidth of the Q-filter is sufficiently large (which will be clarified in the next section).
In the figure, the dotted-block is the plant $P(s)$ with the DOB, and $C(s)$ is the {\it outer-loop controller} that is designed for the nominal model $P_\nn(s)$ (or \eqref{eq:nominal}) of the actual plant $P(s)$.
Since $C(s)$ is designed without considering disturbance and plant's uncertainty, it is the responsibility of the DOB to make the dotted-block behave like $P_\nn(s)$ so that the closed-loop system with $C(s)$ operates as expected.
Then, the design factor is the so-called {\it Q-filters} $Q_A(s)$ and $Q_B(s)$ in the figure.
Conventionally, they are taken as a {\it stable} low-pass filter given by
\begin{equation}\label{eq:Qfilter}
	Q_A(s) = Q_B(s) = \frac{c_{\mu-1} (\tau s)^{\mu-1} + \cdots + c_1 (\tau s) + c_0}{(\tau s)^\mu + a_{\mu-1} (\tau s)^{\mu-1} + \cdots + a_0}
\end{equation}
where $\mu \ge \nu$ and $\tau$ is a positive constant that determines the bandwidth.
As $\tau$ gets smaller, the bandwidth of this filter becomes larger.\footnote{It is trivially verified that the Q-filters have its pole at $\lambda/\tau$ with $\lambda$ being a root of $s^\mu+a_{\mu-1}s^{\mu-1}+\cdots+a_0$.	So the bandwidth is proportional to $1/\tau$.}
We take $c_0 = a_0$ (in order to have the dc-gain one) and $c_{\mu - 1} = c_{\mu-2} = \cdots = c_{\mu-(\nu-1)} = 0$ so that the relative degree of the Q-filter is at least $\nu$.
The latter property is required in order to implement the block $P_\nn^{-1}(s)Q_B(s)$ in the figure, which then becomes a proper transfer function together.
So, the design task is to choose $\tau$, $a_i$, and the remaining $c_i$ appropriately for the control goal that the DOB with $C(s)$ robustly stabilizes the plant (against the uncertainty of the plant) and rejects the effect of the external disturbance on the output $y$ (so that the nominal performance is recovered).

\section{A Closer Look at Conventional DOB}\label{sec:2.2}

For simplicity of presentation, let us consider, from now on, a general example of a third order uncertain plant with relative degree $\nu = 2$ (i.e., $n=3$ and $m=1$) in \eqref{eq:P(s)}, and its nominal model
$$P_\nn(s) = \frac{\beta_{\nn,1} s + \beta_{\nn,0}}{s^3 + \alpha_{\nn,2} s^2 + \alpha_{\nn,1} s + \alpha_{\nn,0}}, \quad \beta_{\nn,1} \not = 0.$$
This nominal model has the normal form realization (like \eqref{eq:sys}) with $g_\nn = \beta_{\nn,1}$, $S_\nn = -\beta_{\nn,0}/\beta_{\nn,1}$, $\phi_{\nn,2} = \beta_{\nn,0}/\beta_{\nn,1} - \alpha_{\nn,2}$, $\phi_{\nn,1} = - (\beta_{\nn,0}/\beta_{\nn,1}) \phi_{\nn,2} - \alpha_{\nn,1}$, $\psi_\nn = - (\beta_{\nn,0}/\beta_{\nn,1}) \phi_{\nn,1} - \alpha_{\nn,0}$, and $G_\nn = [G_{\nn,1},G_{\nn,2}]=[1,0]$.
Suppose that the Q-filter has the same relative degree as $P_\nn(s)$ and is taken as
\begin{equation}\label{eq:QAB}
	Q_A(s) = Q_B(s) = \frac{a_0/\tau^2}{s^2 + (a_1/\tau) s + (a_0/\tau^2)}.
\end{equation}
A realization of the filter $Q_A(s)$ (in Fig.~\ref{fig:CDOB}) is obtained by
\begin{align}\label{eq:p}
	\begin{split}
		\begin{bmatrix} \dot p_1 \\ \dot p_2 \end{bmatrix} &= \begin{bmatrix} 0 & 1 \\ -\frac{a_0}{\tau^2} & - \frac{a_1}{\tau} \end{bmatrix} \begin{bmatrix} p_1 \\ p_2 \end{bmatrix} + \begin{bmatrix} 0 \\ \frac{a_0}{\tau^2} \end{bmatrix} u \\
		y_p &= p_1.
	\end{split}
\end{align}
On the other hand, the transfer function 
$$P_\nn^{-1}(s)Q_B(s) = \frac{(s^3 + \alpha_{\nn,2} s^2 + \alpha_{\nn,1} s + \alpha_{\nn,0})(a_0/\tau^2)}{(\beta_{\nn,1} s + \beta_{\nn,0})(s^2 + (a_1/\tau) s + (a_0/\tau^2))}$$ 
in Fig.~\ref{fig:CDOB} can be realized as\footnote{
Assuming that $P_\nn(s)$ is realized as in \eqref{eq:nominal} with the input $\hat u_\nn$ and the output $v$, the inverse $P_\nn^{-1}(s)$ can be written as (because $x_1$ is the output and $x_2$ is the derivative of the output)
$$\dot z_\nn = S_\nn z_\nn + G_{\nn,1} v, \qquad \hat u_\nn = \frac{1}{g_\nn} (\ddot{v} - \phi_{\nn,1} v - \phi_{\nn,2} \dot{v} - \psi_\nn z_\nn).$$
Now, since $Q_B(s)$ (assuming the input is $y$ and the output is $q_1$) can be realized as $\dot q_1 = q_2$, $\dot q_2 = -\frac{a_0}{\tau^2} q_1 - \frac{a_1}{\tau} q_2 + \frac{a_0}{\tau^2} y$. 
Then, $P_\nn^{-1}(s)Q_B(s)$ can be written as \eqref{eq:zq} since $v = q_1$, $\dot v = q_2$, and $\ddot v = \dot q_2$.
}
\begin{align}
	&\begin{bmatrix} \dot z_\nn \\ \dot q_1 \\ \dot q_2 \end{bmatrix} = \begin{bmatrix} S_\nn & G_{\nn,1} & G_{\nn,2} \\ 0 & 0 & 1 \\ 0 & -\frac{a_0}{\tau^2} & -\frac{a_1}{\tau} \end{bmatrix} \begin{bmatrix} z_\nn \\ q_1 \\ q_2 \end{bmatrix} + \begin{bmatrix} 0 \\ 0 \\ \frac{a_0}{\tau^2} \end{bmatrix} y , \label{eq:zq} \\
	&\; \; \hat u_\nn = \begin{bmatrix} -\frac{\psi_\nn}{g_\nn}, \; -\frac{1}{g_\nn}(\phi_{\nn,1} + \frac{a_0}{\tau^2}), \; -\frac{1}{g_\nn}(\phi_{\nn,2}+\frac{a_1}{\tau}) \end{bmatrix} \begin{bmatrix} z_\nn \\ q_1 \\ q_2 \end{bmatrix} \notag\\
	&\qquad\quad + \frac{1}{g_\nn}\frac{a_0}{\tau^2} y . \notag
\end{align}
See \cite{SJ07} for more detailed derivation.
It is noted that $\hat u_\nn$ can be equivalently rewritten in a simpler way as
\begin{align*}
\hat u_\nn &= \frac{1}{g_\nn} \left( -\left( \psi_\nn z_\nn + \phi_\nn q \right) - \left(\frac{a_0}{\tau^2} q_1 + \frac{a_1}{\tau} q_2 - \frac{a_0}{\tau^2} y \right) \right) \\
&= \frac{1}{g_\nn}\left( \dot q_2 - f_\nn(q,z_\nn)\right)
\end{align*}
where $\dot q_2$ is just a shorthand of $-\frac{a_0}{\tau^2}q_1 - \frac{a_1}{\tau}q_2 + \frac{a_0}{\tau^2}y$ (which is motivated by the last row of \eqref{eq:zq}).
Finally, from Fig.~\ref{fig:CDOB}, the input $u$ is given by
\begin{equation}\label{eq:u}
	u = \bar u + y_p - \hat u_\nn = \bar u + p_1 - \frac{1}{g_\nn}\left( \dot q_2 - f_\nn(q,z_\nn)\right).
\end{equation}
With this, we claim two findings.

\subsection{Robust observer is embedded in $P_\nn^{-1}(s)Q_B(s)$.}

It is seen that equation \eqref{eq:zq} is a cascade of two subsystems
\begin{equation}\label{eq:zwithq}
	\dot z_\nn = S_\nn z_\nn + G_\nn q = h_\nn(q,z_\nn)
\end{equation}
and
\begin{equation}\label{eq:q}
	\begin{bmatrix} \dot q_1 \\ \dot q_2 \end{bmatrix} = \begin{bmatrix} 0 & 1 \\ -\frac{a_0}{\tau^2} & - \frac{a_1}{\tau} \end{bmatrix} \begin{bmatrix} q_1 \\ q_2 \end{bmatrix} + \begin{bmatrix} 0 \\ \frac{a_0}{\tau^2} \end{bmatrix} y.
\end{equation}
Since it is not yet clear to see an observer in it, let us transform $(q_1,q_2)$ into $(\bar q_1, \bar q_2) := (q_1 + (a_1/a_0)\tau q_2, q_2)$.
A simple computation yields
\begin{align}
	\begin{bmatrix} \dot {\bar q}_1 \\ \dot {\bar q}_2 \end{bmatrix} & =  \begin{bmatrix} -\frac{a_1}{\tau} & 1 \\ -\frac{a_0}{\tau^2} & 0 \end{bmatrix} \begin{bmatrix} \bar q_1 \\ \bar q_2 \end{bmatrix} 
	+ \begin{bmatrix} \frac{a_1}{\tau} \\ \frac{a_0}{\tau^2} \end{bmatrix} y\notag\\
	& = \begin{bmatrix} 0 & 1 \\ 0 & 0 \end{bmatrix} \begin{bmatrix} \bar q_1 \\ \bar q_2 \end{bmatrix} 
	+ \begin{bmatrix} \frac{a_1}{\tau} \\ \frac{a_0}{\tau^2} \end{bmatrix} (y-\bar q_1).\label{eq:q,HGO}
\end{align}
This is the very form of the {\it high-gain robust observer} for $x$ (but not for $z$), studied in \cite{KP14} and many others.
According to \cite{KP14}, the state $\bar q(t)$ approaches close to $x(t)\in\R^2$ of \eqref{eq:sys} when $\tau$ is sufficiently small (so, the name `high-gain' follows, as seen in \eqref{eq:q,HGO}).
This is true even though the observer has no information about the system \eqref{eq:sys} (so, it is a {\it robust} observer), and the residual error between $\bar q(t)$ and $x(t)$ (when $t$ tends to infinity) becomes smaller as $\tau$ gets smaller \cite{KP14}.
Moreover, from the transformation between $\bar q$ and $q$, we see that $(q(t)-\bar q(t)) \to 0$ as $\tau \to 0$.
This is the underlying reason why $q(t)$ can be used as the estimate of $x(t)$ when the bandwidth of the Q-filter is large (i.e., $\tau$ is small).
Once $q(t)$ is used instead of $x(t)$, we revisit \eqref{eq:zwithq} and see that it is nothing but \eqref{eq:desired} with $x$ replaced with its estimate.
Verily, the nominal $z$-subsystem \eqref{eq:desired} is implemented in the part of controller, $P_\nn^{-1}(s)Q_B(s)$.
So it can be said that the role of $P_\nn^{-1}(s)Q_B(s)$ is to construct the nominal $z$-subsystem and yield the state estimate of $x$.

\subsection{Estimation of $u_{\rm desired}$ is performed by $Q_A(s)$.}

In the previous subsection, it is seen that $q(t) \to x(t)$ approximately.
Now, in order to have the property $u(t) \to u_{\rm desired}(t)$, we expect from the equation \eqref{eq:u} that
$$p_1(t) \to -\bar u(t) + \frac{1}{g_\nn} (\dot q_2(t) - f_\nn(q(t),z_\nn(t))) + u_{\rm desired}(t).$$
We claim that, if $\tau$ is small and $a_i$ and $c_i$ are suitably chosen, then this is the case.

To see this, let us analyze the (inner) closed-loop system \eqref{eq:p} and \eqref{eq:u} (with $\bar u$, $q$, $y$, and $z_\nn$ viewed as external inputs).
Before going into details, it should be mentioned that the analysis of the behavior of $p_1(t)$, or the system \eqref{eq:p} and \eqref{eq:u} is not very trivial because the $p_1$ term inside $u$ of \eqref{eq:u} cancels the $(2,1)$-element of the $2 \times 2$ system matrix in \eqref{eq:p} so that the stability of the system \eqref{eq:p} with \eqref{eq:u} is seemingly lost (i.e., the system matrix of \eqref{eq:p} does not look like being stable).
Stability is in fact not lost because the (negative) feedback effect of $p_1$ is provided through the state $q$.
To see this, we employ another coordinate change from $(p_1,p_2)$ into $(\pp_1,\pp_2)$ by $\pp_1 := p_1 - \frac{1}{g_\nn}\dot q_2$ and $\pp_2 := p_2 - \frac{1}{g_\nn}\ddot q_2$.
Indeed, a tedious calculation leads to\footnote{
During the calculation, one may note that $\dot \pp_2 = -\frac{a_1}{\tau}\pp_2 + \frac{a_0}{\tau^2}(\bar u + \frac{f_\nn}{g_\nn} - \frac{1}{g_\nn}(f+gu+gd))$ in which $u = \bar u + p_1 - \frac{1}{g_\nn}\dot q_2 + \frac{f_\nn}{g_\nn} = \bar u + \pp_1 + \frac{f_\nn}{g_\nn}$.}
\begin{align}
	&\begin{bmatrix} \dot \pp_1 \\ \dot \pp_2 \end{bmatrix} = \begin{bmatrix} 0 & 1 \\ -\frac{g}{g_\nn}\frac{a_0}{\tau^2} & -\frac{a_1}{\tau} \end{bmatrix} \begin{bmatrix} \pp_1 \\ \pp_2 \end{bmatrix} \label{eq:pp} \\
	& + \begin{bmatrix} 0 \\ \frac{a_0}{\tau^2} \end{bmatrix} \Big\{ (1-\frac{g}{g_\nn})(\bar u + \frac{f_\nn(q,z_\nn)}{g_\nn}) - \frac{1}{g_\nn}(f(x,z)+gd) \Big\}  \notag
\end{align}
in which, the (negative) $(2,1)$-element of the system matrix appears again.
Assume, for the time being, that this system is stable and that the input term (the braced term $\{\cdots\}$ of \eqref{eq:pp}) is constant.
Then, we can conclude that $\pp_2(t) \to 0$ and $\pp_1(t) \to (g_\nn/g) \times (\text{the braced term $\{\cdots\}$ of \eqref{eq:pp}})$.
By plugging this into \eqref{eq:u} (with $p_1 = \pp_1 + \frac{1}{g_\nn}\dot q_2$), one can easily verify that $u(t) \to u_{\rm desired}(t)$.

The issue here is that the input term is not constant but a time-varying signal.
Nevertheless, if $\pp$-dynamics is much faster than this input signal, then the assumption of constant input holds approximately in the relatively fast time scale, and $\pp_1(t)$ quickly converges to its desired (time-varying) value approximately.
This approximation becomes more and more accurate as $\pp$-dynamics gets faster. (See \cite[Sec.~9.6]{Khalil02} for rigorous treatment of this statement.)
A way to make $\pp$-dynamics faster is to take smaller $\tau$, which is seen from the location of eigenvalues.
(One may argue that, by taking smaller $\tau$, the evolution of the high-gain observer state $q(t)$ gets also faster that is contained in the input term. 
While this is true, the state $q(t)$ quickly converges to the relatively slow $x(t)$, and after that, the input term becomes relatively slowly varying.\footnote{The argument here is not very rigorous, but just delivers underlying intuition. See \cite{SJ07,BS08} for more precise proofs using the {\it singular perturbation theory} \cite{Khalil02}.})

Finally, let us inspect whether the system \eqref{eq:pp} is actually stable.
Its characteristic equation is $s^2 + \frac{a_1}{\tau}s + \frac{g}{g_\nn}\frac{a_0}{\tau^2}$ whose roots are $1/\tau$ times the roots of $s^2 + a_1 s + \frac{g}{g_\nn}a_0 =: \ppp_\fff(s)$, and thus, stability of \eqref{eq:pp} is determined by the polynomial $\ppp_\fff(s)$.
Since $g \not = 0$ and the sign of $g$ is known, by letting the nominal value $g_\nn$ have the same sign, the polynomial $\ppp_\fff(s)$ is Hurwitz (because $a_0 > 0$ and $a_1 > 0$ from the stability of the Q-filter).
This is simple because the Q-filter (and thus, the polynomial $\ppp_\fff(s)$) is just of second order.
However, if a higher order Q-filter, like \eqref{eq:Qfilter}, is employed, then $\ppp_\fff(s)$ becomes more complicated (see \cite{SJ09}) as
\begin{multline}\label{eq:pf}
	\ppp_\fff(s) = s^\mu + \left(a_{\mu-1} + \frac{g-g_\nn}{g_\nn}c_{\mu-1}\right) s^{\mu-1} + \cdots \\
	+ \left(a_1 + \frac{g-g_\nn}{g_\nn}c_1\right) s + \left(a_{0} + \frac{g-g_\nn}{g_\nn}c_0\right).
\end{multline}
Then it is not straightforward to ensure that $\ppp_\fff(s)$ is Hurwitz for all variation of $g$, and for this, the coefficients $a_i$ and $c_i$ should be carefully designed.
This observation has been made in \cite{SJ07,BS08,SJ09}.
Fortunately, a design procedure of $a_i$ and $c_i$ has been developed which makes $\ppp_\fff(s)$ remain Hurwitz for arbitrarily large variation of $g$ as long as the upper and lower bounds of $g$ are known. 
See \cite{PJSB12,JPBS14} for the general case, but, for making this paper self-contained, we now quote from \cite{BS08,SJ09} how to choose $a_i$ when $\mu=\nu$, $c_0=a_0$, and $c_i = 0$ for $i=1,\cdots,\mu-1$ (this selection in fact applies to any plant having relative degree $\nu$).
In this case, we have $\ppp_\fff(s) = s^\nu + a_{\nu-1} s^{\nu-1} + \cdots + a_1 s + (g/g_\nn) a_0$.
First, choose $a_{\nu-1}, \cdots, a_1$ so that $\rho(s) := s^{\nu-1} + a_{\nu-1}s^{\nu-2} + \cdots + a_1$ is Hurwitz.
Then, find $\bar k > 0$ such that $s\rho(s) + k = s^{\nu} + a_{\nu-1}s^{\nu-1} + \cdots + a_1s + k$ is Hurwitz for all $0 < k \le \bar k$.
Such $\bar k$ always exists.
Indeed, consider the root locus of the transfer function $1/(s\rho(s))$ with the gain parameter $k$.
Since the root locus includes all points in the complex plane along the real axis to the left of an odd number of poles and zeros (from the right) of the transfer function, and since $1/(s\rho(s))$ has no zeros and has all poles in the left-half plane except one at the origin, the root locus starting at the origin moves to the left as the gain $k$ increases {\it a little} from zero, while the others remain in the open left-half plane for the {\it small} variation of $k$ from zero.
The closed-loop of the transfer function $1/(s\rho(s))$ and the gain $k$ has its characteristic equation $s\rho(s)+k$, and this implies the existence of (possibly small) $\bar k > 0$.
With such $\bar k$ at hand, now choose $a_0 = \bar k/\max\{g/g_\nn\}$ where the maximum is known while $g$ is uncertain.
(So, $a_0$ often tends to be small.)
For the general case, this idea is repeatedly applied.

At last, we note that, if the variation of $g$ is small so that $g \approx g_\nn$, then the term $g-g_\nn$ may be almost zero so that $\ppp_\fff(s)$ remains Hurwitz for all $g$ since the Q-filter is stable so that $s^\mu + a_{\mu-1}s^{\mu-1} + \cdots + a_0$ itself is Hurwitz.
Therefore, with small uncertainties, the stability issue of $\ppp_\fff(s)$ does not stand out and $\ppp_\fff(s)$ is automatically Hurwitz.

\section{Robust Stability of DOB-based Control System}

From the discussions so far, we know that, with Hurwitz $\ppp_\fff(s)$ and small $\tau$, the high-gain observer subsystem \eqref{eq:q} or \eqref{eq:q,HGO} and the subsystem \eqref{eq:pp} are stable.
It is however not enough for the stability of the overall closed-loop system, and let us take the outer-loop controller $C(s)$ into account as well.
Inspecting robust stability of the overall system with DOB is indeed not an easy task in general.
To see the extent of difficulty, let us express $P(s) = N(s)/D(s)$, $P_\nn(s) = N_\nn(s)/D_\nn(s)$, $Q_A(s) = N_\aaa(s)/D_\aaa(s)$, $Q_B(s) = N_\bb(s)/D_\bb(s)$, and $C(s) = N_\ccc(s)/D_\ccc(s)$ where all `$N$' and `$D$' stand for numerator and denominator coprime polynomials, respectively.
Then, the overall system in Fig.~\ref{fig:CDOB} is stable if and only if the characteristic polynomial (we omit `$(s)$' for convenience)
\begin{equation}\label{eq:delta}
	N(N_\nn N_\ccc D_\bb + D_\nn D_\ccc N_\bb) D_\aaa + N_\nn D D_\ccc D_\bb (D_\aaa - N_\aaa)
\end{equation}
is Hurwitz (if there is no unstable pole-zero cancellation between $P_\nn^{-1}(s)$ and $Q_B(s)$, or $P_\nn(s)$ is of minimum phase) \cite{SJ09}.
It is noted that the polynomials $N(s)$ and $D(s)$ are uncertain and so can vary.
Hence, $Q_A$ and $Q_B$ need to be designed such that \eqref{eq:delta} remains Hurwitz under all variations of $N(s)$ and $D(s)$.
This may sound very challenging, but with large bandwidth of Q-filters, it has been proved in \cite{SJ09,JPS14} that the roots of \eqref{eq:delta} form two separate groups, which helps dealing with this challenge.
One group of roots approaches, as $\tau \to 0$, the roots of 
\begin{equation}\label{eq:A1}
	N(s) (D_\nn(s) D_\ccc(s) + N_\nn(s) N_\ccc(s)),
\end{equation}
and the other group of roots approaches $1/\tau$ times the roots of
\begin{equation}\label{eq:A2}
	D^1_\bb(s) ( D^1_\aaa(s) + \gamma N^1_\aaa(s) )
\end{equation}
where $\gamma$ represents $(g-g_\nn)/g_\nn$, which can be simply written as $\gamma = \lim_{s \to \infty} P(s)P_\nn^{-1}(s)-1$ (recalling that $g=\beta_m$ and $g_\nn = \beta_{\nn,m}$).
The superscript `1' in \eqref{eq:A2} implies that $\tau$ is set to 1; for example, $D^1_\bb(s)$ is the denominator $D_\bb(s)$ of $Q_\bb(s)$ with $(\tau s) = s$.
Note that the polynomial \eqref{eq:A2} is nothing but the product of $\ppp_\fff(s)$ in \eqref{eq:pf} and the Hurwitz polynomial $D^1_\bb(s)$.

Since the roots of the characteristic polynomial \eqref{eq:delta} approach the roots of \eqref{eq:A1} and $1/\tau$ times the roots of \eqref{eq:A2} as $\tau \to 0$, robust stability of the overall feedback system is guaranteed if 
\begin{enumerate}
	\item [(A)] $N(s)$ is Hurwitz (i.e., the $z$-subsystem of \eqref{eq:sys} is stable), 
	\item [(B)] $D_\nn D_\ccc+N_\nn N_\ccc$ is Hurwitz (i.e., $C(s)$ internally stabilizes $P_\nn(s)$ (not $P(s)$)),
	\item [(C)] $\ppp_\fff(s)$ remains Hurwitz for all variations of $g$ ($=\beta_m$)
\end{enumerate}
and if the bandwidth of Q-filter is sufficiently large (i.e., $\tau$ is sufficiently small).
If any root of \eqref{eq:A1} or \eqref{eq:A2} appears in the open right-half complex plane, then the overall system becomes unstable with large bandwidth of Q-filter.
Therefore, the above conditions (A), (B), and (C) are necessary and sufficient for robust stability of DOB-based control systems under sufficiently large bandwidth of Q-filters, except the case when any root of \eqref{eq:A1} or \eqref{eq:A2} has zero real part because, in this case, it is not clear in which direction the roots of \eqref{eq:delta} approach the roots of \eqref{eq:A1} and $1/\tau$ times the roots of \eqref{eq:A2}.

It is again emphasized that one group of roots have more and more negative real parts as $\tau \to 0$ (when \eqref{eq:A2} is Hurwitz), and this confirms that some part inside the overall system operates faster than other parts.
This observation goes along with the previous discussions in the state-space (Section 4).
Another way to appreciate \eqref{eq:A1} and \eqref{eq:A2} is the following.
The polynomial \eqref{eq:A2} corresponds to the dynamics which governs the behavior that $u(t) \to u_{\rm desired}(t)$ while \eqref{eq:A1} determines the behavior when $u(t) = u_{\rm desired}(t)$.
Indeed, when $u(t) = u_{\rm desired}(t)$, the $z$-subsystem becomes stand-alone and the uncertain terms $f$ and $g$ are replaced with $f_\nn$ and $g_\nn$ so that stability of $z$-subsystem (or $N(s)$) and stability of the nominal closed-loop (or $D_\nn(s) D_\ccc(s) + N_\nn(s) N_\ccc(s)$) are required.

\section{Robust Transient Response}\label{sec:2.3}

For some industrial applications, {\it robust transient response} (in addition to robust steady-state response) is very important.
For example, if a controller has been designed to satisfy some time-domain specifications such as rising time, overshoot, and settling time for a nominal plant model, then it is desired that the same transient performance is maintained for the real plant under disturbances and uncertainties.
By `robust transient response,' we mean that the output trajectory $y(t)$ of the real plant remains close to the output $y_{\rm nominal}(t)$ {\it for all $t \ge 0$} (i.e., from the initial time) under disturbances and uncertainties, where $y_{\rm nominal}(t)$ is supposed to be the output of the nominal closed-loop system (with the same initial condition).
How can we achieve robust transient response?
In order for $y(t)$ of \eqref{eq:sys} to be the same as $y_{\rm nominal}(t)$ for all $t \ge 0$, we have to have $y^{(i)}(t) = y^{(i)}_{\rm nominal}(t)$ for $i=0,1,\cdots,\nu$.
This task is achieved if $u(t)=u_{\rm desired}(t)$ for all $t \ge 0$,\footnote{If $u(t)=u_{\rm desired}(t)$ from the initial time $t=0$, then the outer-loop controller $C(s)$ feels as if the initial condition of the plant is $(x(0),z_\nn(0))$ (not $(x(0),z(0))$). To see this, refer to \eqref{eq:clsys1}--\eqref{eq:clsys4}.} So, the signal $y_{\rm nominal}(t)$ should be understood as the nominal output resulted by the interaction between $C(s)$ and the nominal plant \eqref{eq:clsys1}--\eqref{eq:clsys3}. 
which is actually the action required for the DOB.\footnote{
	There are two more approaches in the literature to achieve robust transient response (while the underlying principle that $u(t) \to u_{\rm desired}(t)$ quickly is the same).
	One is the universal controller of \cite{FB08} with high-gain observer. 
	This is much similar to DOB, but there is no inner loop of $Q_A(s)$ in Fig.~\ref{fig:CDOB} (which yields $1/(1-Q_A(s))$). Instead, its role is played by a large static gain.
	The other one is the $L_1$ adaptive control of \cite{KKH11,CH07}, for which a constructive design method of the controller} still lacks though.
In fact, this is another reason why $u(t)$ should converge {\it quickly} to $u_{\rm desired}(t)$.
It is indeed because, if $u(t)$ converges to $u_{\rm desired}(t)$ rather slowly, then the time interval for $y^{(\nu)}(t) \not \approx y^{(\nu)}_{\rm nominal}(t)$ becomes large and so, even after $u(t)$ converges to $u_{\rm desired}(t)$, two signals $y(t)$ and $y_{\rm nominal}(t)$ are already different and so afterward.

A remaining problem is that, while making the convergence $u(t) \to u_{\rm desired}(t)$ faster, $u(t)$ may incur very large over/undershoot before it gets close to $u_{\rm desired}(t)$, and this large excursion makes robust transient of $y(t)$ much difficult because large difference between $y^{(\nu)}(t)$ and $y^{(\nu)}_{\rm nominal}(t)$ for short time period may be enough to hamper the property $y(t) \approx y_{\rm nominal}(t)$ during and even after the short time period.
(The situation is related to the peaking phenomenon, which has been studied in \cite{SK91}, of the high-gain observer that is embedded in the $P_\nn^{-1}(s)Q_B(s)$ block.)
A well-known remedy is to insert a saturation element in the feedback loop \cite{EK92} in order to prevent the large excursion of the control signal $u(t)$ from entering the plant.
This technique has been taken for the DOB structure in \cite{BS08,BS09} as in Fig.~\ref{fig:MDOB}.
By the saturation element, the closed-loop system loses `global' stability and the region of attraction in the state-space is restricted.
But, by taking the inactive range of the saturation element sufficiently large, one can secure arbitrarily large region of attraction (which is so-called `semi-global' stabilization that is often enough in practice).
In fact, the saturation should not become active during the steady-state operation of the control system.
In the DOB based control system, it becomes active just for short transient period when $u(t)$ experiences unnecessarily large excursion due to small $\tau$.
See \cite{BS08,BS09} for more details.

	\begin{center}
		\includegraphics[width=0.4\textwidth]{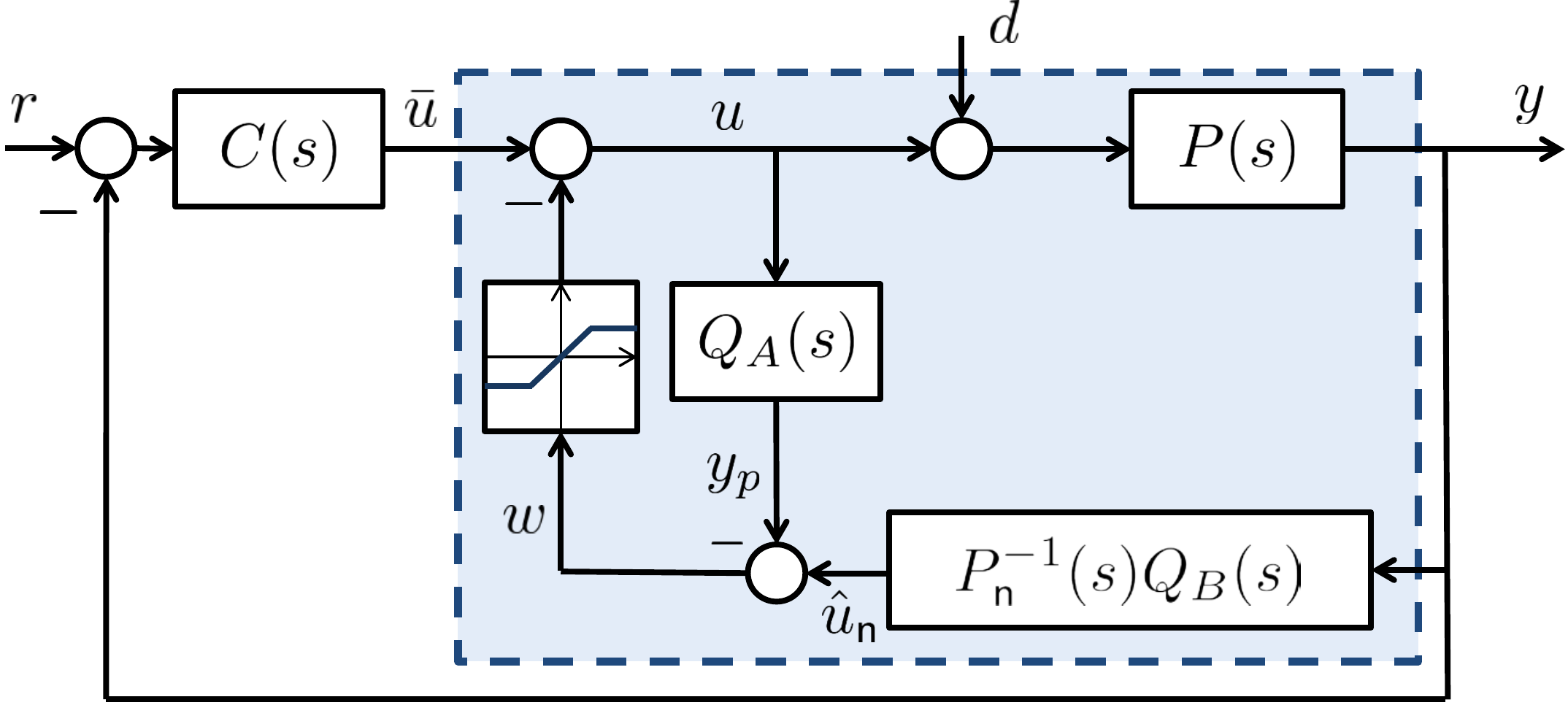}
		\figurecaption{DOB structure with saturation for robust transient response}\label{fig:MDOB}
	\end{center}

\section{Extensions}

The analysis of this paper allows more extensions as follows.
\begin{itemize}
	\item Nonlinear plant: The analysis of the previous sections also applies to single-input-single-output nonlinear plants as long as they have well-defined relative degree so that the plant can be represented as in the normal form like \eqref{eq:sys}.\footnote{See \cite{Isidori95} about how to transform a nonlinear system with well-defined relative degree into the normal form.}
	See \cite{BS08} for more details.
	
	\item MIMO plant: Multi-input-multi-output plants also admit the DOB. See \cite{BS09} for details.
	
	\item Reduced-order DOB: Since the same Q-filters are found in Fig.~\ref{fig:CDOB}, they could be merged into one in order to reduce the dimension of the DOB.
	See \cite{BS11a,BS14}.
	
	\item Exact (not approximate) rejection of disturbances: Nominal performance recovery (or disturbance rejection) studied in this paper is based on the convergence $u\rightarrow u_{\rm desired}$.
	This convergence, however, is inherently approximate because estimation of $u_{\rm desired}(t)$ is approximate and the convergence of $u(t)$ to the estimate of $u_{\rm desired}(t)$ is also approximate, although the approximation becomes more and more accurate as $\tau$ gets smaller.
	However, if the disturbance is generated by a known generating model (which is called an {\it exosystem}), then exact rejection of the disturbance, without relying on smallness of $\tau$, is possible.
	(One typical example is the sinusoidal disturbance with known frequency.)
	The tool used for this purpose is the well-known {\it internal model principle} \cite{FW76}, and the controller design has been studied under the name of {\it output regulation} in, e.g., \cite{Huang04}.
	The DOB (more specifically, the Q-filters) can be modified to include the generating model of the disturbance, so that the internal model principle holds for the closed-loop system.
	The initial result in this direction is found in \cite{YKIH96}. 
	For embedding the generating model of sinusoidal disturbance with robust stabilization, refer to \cite{PJSB12,JPBS14,PJS14}.

	\item Different Q-filters: The conventional DOB consists of two same Q-filters, while their roles are inherently different as discovered in Section 4. Such observation has triggered subsequent works \cite{JPS14,PJS14,Han13} in which the Q-filters are refined separately for certain purposes. For instance, since the estimation of $u_{\rm desired}(t)$ is mainly performed by $Q_A(s)$, it is sufficient to embed the generating model of disturbance just into $Q_A(s)$ (but not into $Q_B(s)$) for the exact disturbance rejection \cite{PJS14}. Another example is to use higher order of $Q_B(s)$ (while the order of $Q_A(s)$ being kept the same) in order to have more reduction of the effect of measurement noise \cite{Han13}.

	\item Use of state feedback controller $C$: Since the estimate of the plant's state $x$ is provided by the DOB, the outer-loop controller $C$ can be of state feedback type if the full state is considered as $(x,z_\nn)$ where $z_\nn$ is also provided by the DOB. See \cite{Ha14} for this combination.
		
	\item Input saturation: In practice, it is natural that the control input $u$ is limited. A preliminary result on DOB-based controller under input saturation has been reported in \cite{Yoon15}, where the authors presented an LMI to find the control gain for a state feedback controller and the parameters of DOB at the same time.
	
	\item Discrete-time implementation of DOB: While all the above results are discussed in the continuous-time domain, for implementing the DOB in the digital devices, DOB is constructed in the discrete-time domain.
	At first glance, stability seems to remain guaranteed if the continuous-time DOB is discretized by fast sampling.
	However, fast sampling (with zero-order hold) of a continuous-time system introduces additional zeros (which are called {\it sampling zeros} \cite{AHS84,YG14}), and worse yet, at least one of them is always unstable when the relative degree $\nu \ge 3$.
	This causes another trouble that the sampled-data model of the plant becomes a non-minimum phase system, which seems violating the necessary condition for stability as discussed earlier.
	In \cite{PJLS15}, it has been found that the Q-filters in the discrete-time domain can be specially designed to take care of unstable sampling zeros as well while maintaining all good properties in the continuous-time domain.
	
\end{itemize}

\section{Restrictions}

We have so far looked at the conventional DOB and its extensions, and found that the DOB is a powerful tool for robust control and disturbance rejection.
These benefits came under the requirement that the bandwidth of the Q-filter is sufficiently large, and this in turn imposed a few restrictions as follows.

\begin{itemize}
	
	\item Unmodelled dynamics: If there is unmodelled dynamics in the real plant, then the relative degree $\nu$ of the plant $P(s)$ may be different from that of the nominal model $P_\nn(s)$. 
	In this case, it has been actually reported in \cite{JJS14} that large bandwidth may lead to instability of the overall system, even though a remedy for a few cases is also suggested in \cite{JJS14}. See also \cite{JJS11}.
	
	\item Non-minimum phase plant: If the real plant is of non-minimum phase (i.e., the zero dynamics is unstable), then large bandwidth of Q-filters makes the overall system unstable, as discussed before.
	To overcome this restriction, another structure with different Q-filters has been proposed in \cite{JSS10}, but the problem is still open in general.
	
	\item Sign of high frequency gain: We have seen that the sign of the (uncertain) high-frequency gain $g$ (or $\beta_m$) should be known.
	If it is not the case, then, as a workaround, the Nussbaum gain technique \cite{Nussbaum83} has been employed in \cite{Yoon12}, while more study is called for in this direction. 
	
	\item Measurement noise: Large bandwidth of Q-filters may also yield insufficient reduction of noise at high frequencies while we refer to \cite{Xie10,NS13,Han13} for a possible modification of the DOB structure to enhance the noise reduction capability.
	
\end{itemize}

We emphasize that, if the bandwidth of Q-filter is severely limited by some reason, then the desired steady-state/transient performance and the robust stabilization may not be obtained simply because the analysis so far is no longer valid.
Fortunately, an appropriate bandwidth of Q-filter for robust stabilization and disturbance rejection, varies from system to system, and it turns out that the bandwidth of Q-filter need not be too large in many practical cases.
For instance, a reasonable choice of bandwidth worked in the experiments of \cite{YCS09,BSPS10}.
The analysis in the related papers yields an upper bound for $\tau$ that works, in theory.
But, since it is often too conservative, finding suitable $\tau$ is done usually by repeated simulations or by tuning it in experiments.

On the other hand, there are other approaches that do not explicitly rely on large bandwidth of Q-filter for robust stability  \cite{KK99,GG02,ESC01,CYCWHS03,WT04,SD02,BT99,SO13,SO14}. 
Among them, the most popular one is to employ the tool `small-gain theorem' \cite{KK99,GG02,ESC01}.
However, it gives a sufficient condition for stability and so may yield conservatism.
More specifically, for an uncertain plant with multiplicative uncertainty (i.e., $P(s) = P_{\nn}(s)(1 + \Delta(s))$ where $\Delta(s)$ is an unknown stable transfer function), the condition for robust stability is derived as $\| \Delta \|_{s = j \omega} < \|(1 + P_{\nn}C)/(Q + P_{\nn}C) \|_{s = j \omega}$ for all $\omega$ \cite{ESC01}.
Then, the size of uncertainty is severely limited (considering the typical case $0 \le \|Q\|_{s=j\omega} \le 1$).

Another one is working with the {\it state (not output)} measurements as in \cite{Chen04,GC05,KRK10}.
On the other hand, assuming that the disturbance is an output of a generating model (the exosystem), the disturbance observers which estimate the disturbance by estimating the exosystem's state, are often proposed and combined with many well-established controllers such as sliding mode control or model predictive control. 
See, for example, \cite{LYCWC14,GC13} for more details.

In spite of these restrictions, study of DOB under large bandwidth of Q-filters is worthwhile since it illustrates the role of each component of DOB, and yields useful insights for further study of DOB.
It also shows ideal performance that can be achieved under arbitrarily large parametric uncertainty and disturbances, and constructive design guides of Q-filters are derived.

Finally, we close this tutorial with a disclaimer that the purpose of this tutorial is not to survey exhaustive list of related contributions on DOB, but just to present a new perspective of the authors about the DOB. 
So, some important contributions on DOB may have been omitted.
This tutorial is written in less formal manner, and more rigorous, theorem-proof style, arguments can be found in the references cited in the text.

\bigskip

\appendix
{\bf\noindent Appendix}
\nopagebreak

The conversion from \eqref{eq:origsys} to \eqref{eq:sys} is done by the following procedure.
First, it is noted that $\ccc\AAA^{i-1}\bb = 0$ for $i=1, \cdots, \nu-1$ and $\ccc\AAA^{\nu-1}\bb \not = 0$, which is because the relative degree of $P(s)$ is $\nu=n-m$ so that the input $u$ does not appear explicitly until we take time derivative of $y$ up to $\nu$ times.
(This is again because $P(s)$ remains strictly proper until we multiply it by $s^{\nu}$.)
Then, the row vectors $\ccc\AAA^{i-1}$ for $i=1, \cdots, \nu$ are linearly independent. 
Indeed, it is seen that
\begin{multline*}
	\begin{bmatrix} \ccc \\ \ccc\AAA \\ \vdots \\ \ccc\AAA^{\nu-1} \end{bmatrix} 
	\begin{bmatrix} \bb & \AAA\bb & \cdots & \AAA^{\nu-1}\bb \end{bmatrix} \\
	= \begin{bmatrix} 0 & \cdots & 0 & \ccc\AAA^{\nu-1}\bb \\
		0 & \cdots & \ccc\AAA^{\nu-1}\bb & * \\
		\vdots & \vdots & \vdots & \vdots \\
		\ccc\AAA^{\nu-1}\bb & * & * & * \end{bmatrix}
\end{multline*}
from which, two matrices on the left-hand side have full row/column ranks.
Then, one can easily find a matrix $\Phi \in \R^{m \times n}$ such that 
$$\begin{bmatrix} \ccc \\ \ccc\AAA \\ \vdots \\ \ccc\AAA^{\nu-1} \\ \Phi \end{bmatrix} =: \TTT \; \text{is nonsingular and $\Phi \bb = 0$.}$$
Indeed, the dimension of the left nullspace of $\bb$ is $n-1$, and from the derivation, it is clear that $\ccc\AAA^{\nu-1}$ does not belong to the left nullspace of $\bb$ while $\nu-1$ vectors $\ccc\AAA^{i-1}$, $i=1,\cdots,\nu-1$, belong to it.
So, one can find $(n-1) - (\nu-1) = m$ linearly independent row vectors in the subspace that are linearly independent of $\ccc\AAA^{i-1}$, $i=1,\cdots,\nu-1$.
	
Now, let
$$\bar \EEE = \begin{bmatrix} 0 & 0 & \cdots & 0 \\
\ccc\EEE & 0 & \cdots & 0 \\
\ccc\AAA\EEE & \ccc\EEE & \cdots & 0 \\
\vdots & \vdots & \ddots & 0 \\
\ccc\AAA^{\nu-2}\EEE & \ccc\AAA^{\nu-3}\EEE & \cdots & \ccc\EEE \end{bmatrix}, \quad
\bar \ddd = \begin{bmatrix} \ddd \\ \dot{\ddd} \\ \ddot \ddd \\ \vdots \\ \ddd^{(\nu-2)} \end{bmatrix}
$$
and let
$$\begin{bmatrix} x \\ z \end{bmatrix} = \TTT\xxx + \begin{bmatrix}\bar \EEE \\ 0_{m \times q(\nu-1)} \end{bmatrix} \bar \ddd$$
where $x \in \R^\nu$ and $z \in \R^m$.
For convenience, let $\TTT^{-1} =: [\TTT_a, \TTT_b]$ with $\TTT_a \in \R^{n \times \nu}$ and $\TTT_b \in \R^{n \times m}$.
Then, it can be seen that system \eqref{eq:sys} is obtained. 
Indeed, we verify that 
\begin{align*}
	y &= \ccc \xxx = x_1 \\
	\dot x_1 &= \ccc\AAA\xxx + \ccc\EEE\ddd = x_2 \\
	\dot x_2 &= \ccc\AAA^2\xxx + \ccc\AAA\EEE\ddd + \ccc\EEE\dot\ddd = x_3 \\
	&\vdots \\
	\dot x_\nu &= \ccc\AAA^{\nu}\xxx + \ccc\AAA^{\nu-1}\bb u + \sum_{j=1}^\nu \ccc\AAA^{\nu-j}\EEE\ddd^{(j-1)} \\
	&= \ccc\AAA^{\nu}\TTT^{-1}\left(\begin{bmatrix} x - \bar{\EEE}\bar\ddd \\ z \end{bmatrix}\right) + gu + \sum_{j=1}^\nu \ccc\AAA^{\nu-j}\EEE\ddd^{(j-1)} \\
	&= \phi x + \psi z + g u + g d
\end{align*}
by letting $g = \ccc\AAA^{\nu-1}\bb$,\footnote{
	$g$ is in fact $\beta_m$ in \eqref{eq:P(s)}. Indeed, from \eqref{eq:P(s)}, we have that $s^\nu P(s) = \beta_m + (\text{a strictly proper transfer function})$. This implies that $y^{(\nu)} = \beta_m u + \cdots$, and from this, it is clear that $g = \beta_m$.
} $\phi = \ccc\AAA^\nu \TTT_a$, $\psi = \ccc\AAA^\nu \TTT_b$, and 
\begin{equation}\label{eq:ddd}
	d = \frac1g \left(-\ccc\AAA^\nu \TTT_a\bar\EEE\bar{\ddd} + \sum_{j=1}^\nu \ccc\AAA^{\nu-j}\EEE\ddd^{(j-1)} \right).
\end{equation}
We also verify that 
\begin{align*}
	\dot z &= \Phi \AAA \xxx + \Phi \bb u + \Phi \EEE \ddd \\
	&= \Phi\AAA \TTT^{-1}\left(\begin{bmatrix} x - \bar{\EEE}\bar\ddd \\ z \end{bmatrix}\right) + 0 + \Phi\EEE\ddd \\
	&= Sz + Gx + d_z
\end{align*}
with $S = \Phi\AAA\TTT_b$, $G = \Phi\AAA\TTT_a$, and $d_z := \Phi\EEE\ddd - \Phi\AAA\TTT_a \bar{\EEE}\bar \ddd$.

It is noted that, if there is no disturbance, then the conversion from \eqref{eq:origsys} to \eqref{eq:sys} is nothing but the similarity transformation by the matrix $\TTT$.
Therefore, the poles and zeros of $P(s)$ are preserved both in \eqref{eq:origsys} and \eqref{eq:sys}.
While it is clear that the eigenvalues of the system matrix of \eqref{eq:sys} are the poles, it is important to note that the eigenvalues of $S$ are the zeros of the system.
Indeed, let $\lambda \bar z = S \bar z$ with $\lambda$ and $\bar z$ being the eigenvalue and the eigenvector of $S$, respectively.
Then, the following equality holds:
$$\begin{bmatrix}
\lambda & -1 & 0 & \cdots & 0_{1 \times m} & 0 \\
0 & \lambda & -1 & \cdots & 0_{1 \times m} & 0 \\
\vdots & \vdots & \vdots & \ddots & \vdots \\
& & -\phi & & -\psi & g \\
& & -G & & \lambda I - S & 0 \\
1 & 0 & 0 & \cdots & 0_{1 \times m} & 0
\end{bmatrix} \begin{bmatrix} 0 \\ 0 \\ 0 \\ \vdots \\ \bar z \\ \frac{\psi \bar z}{g} \end{bmatrix} = 0$$
in which, the left matrix is the Rosenbrock system matrix \cite{Chen99} of \eqref{eq:sys}.
The above equation implies that the Rosenbrock matrix loses rank with $\lambda$, and thus, the value $\lambda$ is the zero of the system \cite{Chen99}.

\vspace{.5\baselineskip} {\selectfont\scriptsize
{Hyungbo Shim} received his B.S., M.S., and Ph.D. degrees from Seoul National University, Korea, in 1993, 1995 and 2000, respectively. From 2000 to 2001 he was a post-doctoral researcher at University of California, Santa Barbara. Since 2003, he has been with Seoul National University, where he is now a professor. He has served as Associate Editor for the journals IEEE Trans. on Automatic Control and Automatica.
\par}

\vspace{.5\baselineskip} {\selectfont\scriptsize
{Gyunghoon Park} received his B.S. degree in the School of Electrical and Computer Engineering from Sungkyunkwan University in 2011, and M.S. degree from the School of Electrical Engineering and Computer Science, Seoul National University, in 2013, respectively. Since 2013, he has been working toward his Ph. D. degree at Seoul National University.
\par}

\vspace{.5\baselineskip} {\selectfont\scriptsize
{Youngjun Joo} received his B.S., M.S., and Ph. D. degrees from the School of Electrical Engineering and Computer Science, Seoul National University, Korea, in 2005, 2007, and 2014, respectively. From 2014 to 2015, he was a post-doctoral researcher at Hanyang University, Korea. Since 2015, he has been working as a post-doctoral researcher at University of Central Florida, USA.
\par}

\vspace{.5\baselineskip} {\selectfont\scriptsize
{Juhoon Back} received the B.S. and M.S. degrees in Mechanical Design and Production Engineering from Seoul National University, in 1997 and 1999, respectively. He received the Ph.D. degree from the School of Electrical Engineering and Computer Science, Seoul National University in 2004. From 2005 to 2006, he worked as a research associate at Imperial College London, UK. Since 2008 he has been with Kwangwoon University, Korea, where he is currently an associate professor.
\par}

\vspace{.5\baselineskip} {\selectfont\scriptsize
{Nam H.~Jo} received the B. S.,  M. S., and Ph. D. degrees from School of Electrical Engineering, Seoul National University, Seoul, Korea, in 1992, 1994 and 2000, respectively. From 2001 to 2002, he worked as a Senior Research Engineer at Samsung Electronics, Suwon, Korea. Since 2002, he has been with the Department of Electrical Engineering at Soongsil University, Seoul, Korea, where he is currently a professor.
\par}

\end{multicols}
\end{document}